\documentclass[onecolumn]{elsarticle}

\usepackage{amsmath,amssymb,url}
\usepackage{amsthm}
\usepackage{graphicx,subfigure}
\usepackage[adobe-utopia]{mathdesign}
\usepackage[T1]{fontenc}
\usepackage{braket}
\usepackage{fancyvrb}
\usepackage{color}
\usepackage{physics}

\begin{document}

\title{High Precision Statistical Landau Gauge Lattice Gluon Propagator Computation vs.~the Gribov-Zwanziger approach}

\author{David Dudal$^{a,b}$}
\ead{david.dudal@kuleuven.be}
\address{$^a$KU Leuven Campus Kortrijk -- Kulak, Department of Physics, Etienne Sabbelaan 53 bus 7657, \\ 8500 Kortrijk, Belgium  \\
$^b$Ghent University, Department of Physics and Astronomy, Krijgslaan 281-S9, 9000 Gent, Belgium}

\author{Orlando Oliveira$^{c,d,e}$}
\ead{orlando@fis.uc.pt}

\author{Paulo J.~Silva$^c$}
\ead{psilva@uc.pt}

\address{$^c$Centro de F\'{i}sica da Universidade de Coimbra, Departamento de F\'{i}sica, Universidade de Coimbra, 3004-516 Coimbra, Portugal \\
$^d$Laborat\'orio de F\'{\i}sica Te\'orica e Computacional, Universidade Cruzeiro do Sul, 01506-000 \\ S\~ao Paulo, SP, Brazil \\
$^e$Instituto Tecnol\'ogico de Aeron\'autica, DCTA, 12228-900, S. Jos\'e dos Campos, Brazil}



\date{\today}

\begin{abstract}
In this article we report on lattice results for the Landau gauge gluon propagator using large statistical ensembles. In particular,
the compatibility of the lattice data with the tree level predictions of the Refined Gribov-Zwanziger and the Very Refined Gribov-Zwanziger
actions is analysed, complementing earlier work using small-scale statistics.
Our results show that the data is well described by the tree level estimate only up to momenta $p \lesssim 1$ GeV and clearly favors the so-called Refined Gribov-Zwanziger
scenario propagator. For the Very Refined Gribov-Zwan\-ziger action, this results implies a particular relation between $d=2$ condensates.
Furthermore, we also provide a global fit of the lattice data that interpolates between the above scenario at low momenta and the usual continuum one-loop re\-nor\-ma\-li\-za\-tion
improved perturbation theory after introducing an infrared log-re\-gu\-la\-ri\-zing term.
\end{abstract}

\begin{keyword}
lattice QCD, gluon propagator, Landau gauge, Gribov-Zwanziger quantization.
\end{keyword}

\maketitle

The Landau gauge lattice gluon propagator has been computed both for the SU(2) and SU(3) gauge groups in four dimensions by several groups
using various lattice spacings and physical volumes \cite{Cucchieri:2007md,Cucchieri:2007zm,Bogolubsky:2009dc,Dudal:2010tf,Ilgenfritz:2010gu,Oliveira:2010xc,Cucchieri:2011ig,Oliveira:2012eh,Duarte:2016iko}. These calculations gave a clear picture on the behaviour of the gluon propagator and a reasonable
understanding of the finite space and volume effects on the gluon two point correlation function.
The access to the deep infrared region was achieved by performing simulations on huge physical volumes at relatively small $\beta$ values, i.e.
large lattice spacings, which seems to suppress the propagator at small momenta. These simulations on huge physical volumes, which had in mind the
deep infrared region, have a limited access to the ultraviolet region and provide no information on the propagator for momenta $p \gtrsim 4$ GeV. On
the other hand, the simulations for the propagators suggest that finite volume effects are marginal for simulations on physical volumes $\gtrsim ( 6.5$ fm$)^4$
that use a lattice spacing $a \lesssim 0.1$ fm.

As described in~\cite{Boucaud:2017ksi,Duarte:2017wte} some of the differences in the infrared, if not all, of the results reported for the gluon propagator by the
various research groups can be attributed to a not so precise setting of the lattice scaling.
Indeed it was found that a change of about 3\% on the value for the lattice spacing improve considerably
the agreement  of the lattice data computed at different $\beta$ values.

The lattice gluon propagator has oftentimes been interpreted in terms of the tree level predictions of the Gribov-Zwanziger action and its
generalizations~\cite{Dudal:2010tf,Cucchieri:2011ig,Vandersickel:2012tz,Dudal:2012zx,Oliveira:2012eh,Aouane:2011fv,Bornyakov:2013pha}, with applications in \cite{Fukushima:2012qa,Canfora:2013zna,Canfora:2016xnc}. The Faddeev-Popov construction is plagued with the problem of the Gribov copies which, from a theoretical point of view, invalidates its use to tackle
non-perturbative problems, i.e.~to study the infrared region. The Gribov-Zwanziger action incorporates a restriction of the functional integration space to the first Gribov horizon,
where the eigenvalues of the Faddeev-Popov operator are positive definite, and as such a considerable set of gauge copies is discarded \cite{Gribov:1977wm,Vandersickel:2012tz}.
The Gribov-Zwanziger family of actions do not solve completely the Gribov ambiguity, but at least they provide an improved gauge fixing for the
gauge sector when compared to the Faddeev-Popov generating functional. A complete resolution of the Gribov copies requires implementing the restriction of the functional integration to the fundamental modular region, an open problem with no known solution so far \cite{vanBaal:1991zw}.

The tree level gluon propagator predicted by the Gribov-Zwanziger action and its generalizations is a ratio of polynomials in $p^2$ that,
at large momenta, reduces the Landau gauge propagator to $1/p^2$ and recovers the usual perturbative behaviour for the ultraviolet regime.
In general and for the simulations performed until now, the lattice data for the Landau gauge gluon propagator seems to be compatible with the tree level prediction
of  the so-called Refined Gribov-Zwanziger action. The outcome of the analysis of the lattice data within this picture results in a gluon propagator
that is described by a pair of complex conjugate poles that are emplaced on the negative side of the complex Euclidean $p^2$ plane, namely  Re$(p^2) \approx -250$ MeV and
with $|$Im$(p^2)| \approx 450$ MeV being larger than its real part.
This set of infrared poles configure a non-physical gluon, in the sense that such a particle cannot contribute to the $\mathcal{S}$-matrix. Furthermore,
a propagator with complex conjugate poles rises fundamental questions such as the violation of causality that possibly can be precluded because
the gluon does not seem to be a physically observable, but rather a confined, particle \cite{Stingl:1994nk,Grinstein:2008bg}. Note also that if the gluon propagator is described by a pair of complex conjugate poles, then no K\"all\'{e}n-Lehmann spectral representation can be associated to the gluon. However,  in~\cite{Dudal:2013yva,Oliveira:2016stx}
the authors were able to compute such a spectral representation for the gluon
propagator that describes rather well the lattice data, see also \cite{Strauss:2012dg}. Although these results seem to be contradictory, this is to be expected, in the sense that given a finite set of data points, there is no unique way to associate an analytic continuation of this set to the whole complex plane, and the K\"all\'{e}n-Lehmann representation is doing nothing else than offering a \emph{possible} analytic continuation for a non-physical particle, in contradistinction with the case of a physical propagating degree of freedom, which ought to have a K\"all\'{e}n-Lehmann spectral representation based on general grounds.

The analysis of the lattice data within the Gribov-Zwanziger picture relies on its tree level predictions and, clearly, one should identify corrections that
change its deep infrared, its large momentum region behaviour and that, eventually, might wash out the estimated complex conjugate poles.
It is somehow surprising that tree level perturbation theory within the Gribov-Zwanziger approach is able to
describe so well the Landau gauge lattice gluon propagator or, said it otherwise, that the non-perturbative gluon propagator can be understood from a
perturbative analysis with a conveniently modified action. This seems to be case not only for the Refined Gribov-Zwanziger action, see e.g.~\cite{Dudal:2010tf},
but also using the Curci-Ferrari model, see e.g.~\cite{Tissier:2011ey}.

The analysis till now of the lattice Landau gauge gluon propagator used relatively small ensembles of gauge configurations, with several studies
relying on  $\mathcal{O}(100)$ gauge configurations per value of $\beta$ \cite{Cucchieri:2007md,Cucchieri:2007zm,Bogolubsky:2009dc,Dudal:2010tf,Ilgenfritz:2010gu,Oliveira:2010xc,Cucchieri:2011ig,Oliveira:2012eh,Duarte:2016iko,Leinweber:1998uu,Maas:2017csm}. It is therefore quite natural to wonder whether the compatibility of the Landau gauge lattice gluon propagator and the (tree level) predictions of the
Refined Gribov-Zwanziger action still holds for very large statistical ensembles of gauge configurations, i.e.~is a high statistical precision computation
of the gluon propagator sensitive to the corrections to the tree level result of the Gribov-Zwanziger construction? Of course, such a study also serves as a test of the
predictability of the (Refined and Very Refined)
 Gribov-Zwanziger action. Herein we try to answer these simple questions via a high statistical calculation of the gluon propagator and
presenting dedicated fits. In particular, we will use a minimizing $\chi^2/\text{d.o.f.}$ fitting procedure taking into account correlations between the
different momenta through the covariance matrix. This automatically allows to control the quality of the fit
via the value of the $\nu = \chi^2/\text{d.o.f.}$

In this paper, the Landau gauge gluon propagator is computed for two large ensembles of gauge configurations with large physical volumes,
and the compatibility of the lattice data with the tree level predictions of the family of Gribov-Zwanziger actions, i.e. the Refined
Gribov-Zwanziger (RGZ) and the Very Refined Gribov-Zwanziger (VRGZ) tree level gluon propagators will be investigated.

\section{Lattice Setup, Lattice Data and the Tree Level Gribov-Zwanziger Gluon Propagator}

Herein, we will analyse two large physical volume lattice simulations performed with $\beta = 6.0$, i.e.~with a lattice spacing of $1/a = 1.943$ GeV, i.e.
an $a = 0.1016(25)$ fm, measured from the string tension.
The simulations rely on $64^4$ and $80^4$ lattices whose physical volume are $(6.57$ fm$)^4$ and $(8.21$ fm$)^4$,
respectively. The ensemble associated to the smaller volume uses 2000 gauge configurations rotated to the Landau gauge, while for the largest physical
volume the propagator is computed using 550 gauge configurations rotated to the Landau gauge. The details on the sampling, gauge fixing and definitions
can be found in \cite{Silva:2004bv,Duarte:2016iko}.

In the Landau gauge, the gluon propagator has a single form factor
\begin{equation}
    \langle A^a_\mu (\hat{p}) \, A^b_\nu (\hat{p}^\prime) \rangle = V \, \delta^{ab} \, \delta( \hat{p} + \hat{p}^\prime) \, P_{\mu\nu} ( p ) \, D(p^2) \ ,
\end{equation}
where
\begin{equation}
   \hat{p}_\mu = \frac{ 2 \pi }{a L} \, n_\mu  \, , \qquad n_\mu = 0, \, 1, \, \dots, \, L -1
\end{equation}
is the lattice momentum,
\begin{equation}
  p_\mu = \frac{2}{a} \sin \left( \frac{\pi}{L} n_\mu \right)
\end{equation}
is the continuum momentum,
\begin{equation}
P_{\mu\nu} ( p ) =   \delta_{\mu\nu} - \frac{ p_\mu p_\nu}{p^2}
\end{equation}
the orthogonal projector,  $L$ is the number of lattice points in each space-time direction and $V = L^4$ the lattice volume. The smallest
non-vanishing continuum momentum
accessed in the simulations is 191 MeV for the $64^4$ lattice and 153 MeV for the $80^4$ lattice. For the two simulations the largest physical
continuum momentum
considered is about 7.7 GeV.

The lattice data referred here were renormalised within the MOM scheme at a scale $\mu = 3$ GeV by demanding that
\begin{equation}
  \left.   D( p^2 ) \right|_{p^2 = \mu^2} = \frac{1}{\mu^2} \ .
\end{equation}
This choice of the renormalization point allows to compare results with the previous study performed in~\cite{Dudal:2010tf}.

\begin{figure}[t] 
   \centering
   \includegraphics[width=3.7in]{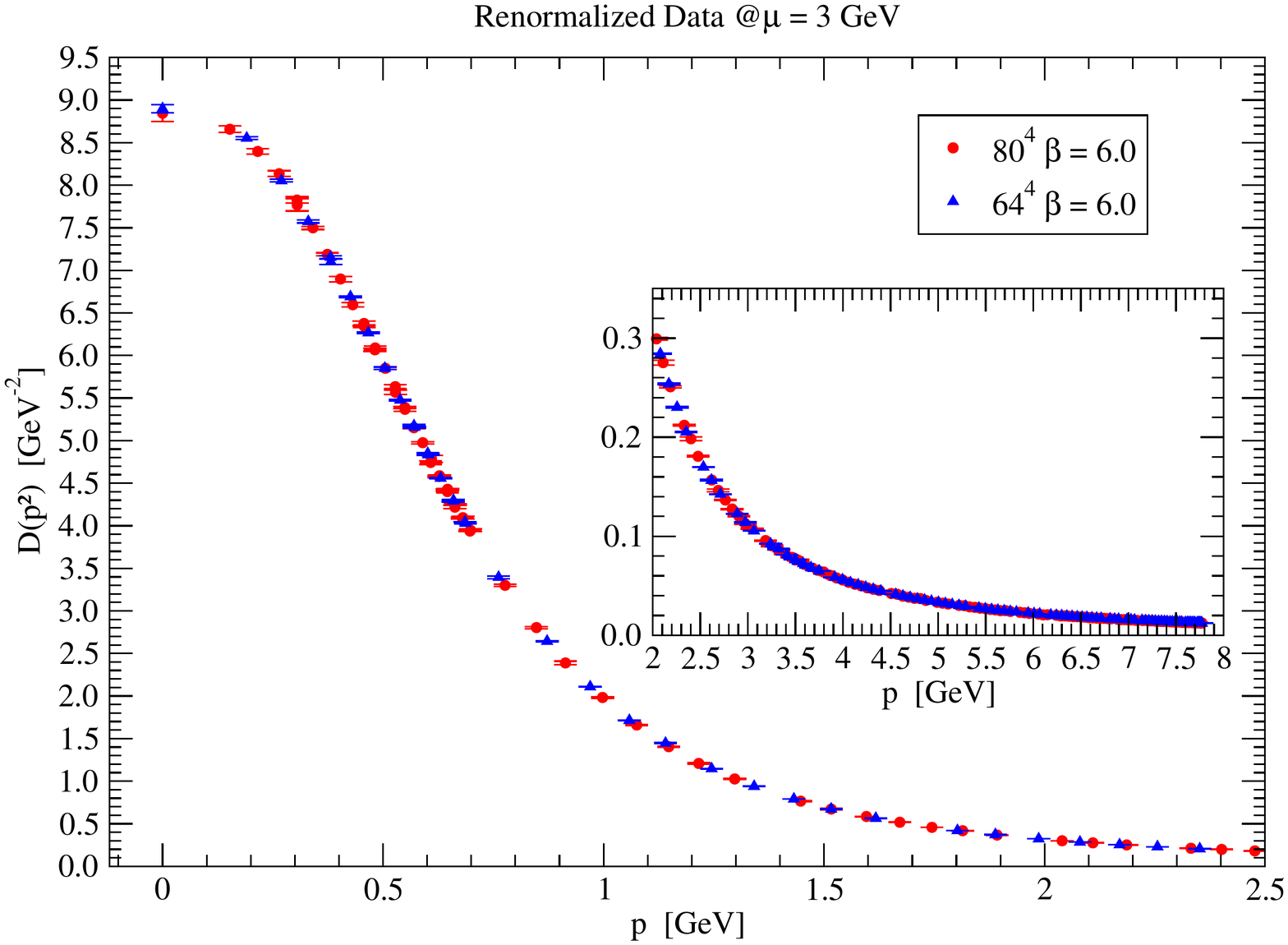} \\
   \includegraphics[width=3.7in]{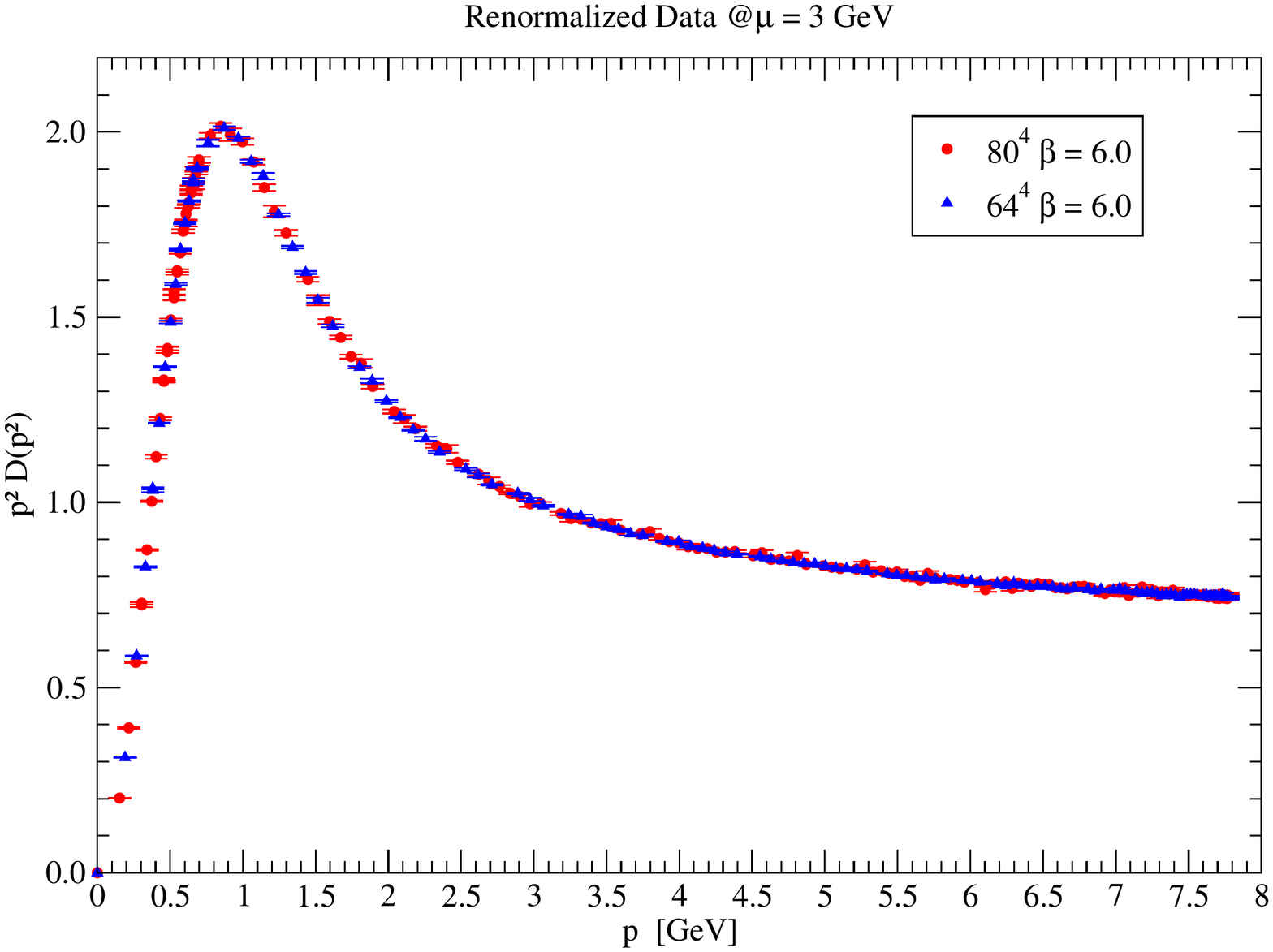}
   \caption{Lattice gluon propagator (top) and dressing function (bottom) data renormalized at $\mu = 3$ GeV for the full range of momenta.}
   \label{fig:gluonprops}
\end{figure}

The renormalization constants were computed by fitting the bare lattice data in the momentum region
$p \in [ 1.5 , 4.5 ]$ GeV for the $64^4$ lattice and $p \in [ 1 , 5 ]$ GeV for the $80^4$ data to the Pad\'e (rational function) approximation
\begin{equation}
   D(p^2) = Z \frac{p^2 + M^2_1}{p^4 + M^2_2 p^2 + M^4_3} \ ,
\end{equation}
where $M_1$, $M_2$ and $M_3$ have mass dimensions, and then rescaling the lattice data using the results of the fit, which is physically motivated by \cite{Dudal:2007cw,Dudal:2008sp,Stingl:1994nk}. The $\nu = \chi^2/\text{d.o.f.}$ of the fits
are 1.43 for the data from the $64^4$ lattice and $0.95$ for the data coming from the simulation using the $80^4$ lattice.
The renormalization constants for the two lattices differ by $0.068\%$. We take this result as an indication that the finite volume effects are negligible
for the fitting range and will thence assume that no systematics are associated with the renormalization procedure.
In the following, we also disregard possible effects due to Gribov copies, in the sense that we did not take into account possible deviations in the generated
Landau gauge ensembles of configurations due to different (local) maxima of the Landau gauge maximizing functional~\cite{Silva:2004bv}.

The problems associated to the definition of the physical scale referred
in~\cite{Boucaud:2017ksi,Duarte:2017wte} are also not taken into account.
Given that we are using a unique $\beta$ value in the simulations, there are no ambiguities coming from the scale setting.

In order to reduce lattice artifacts for momenta above 0.7 GeV, we performed the conical and cylindrical cuts as described in~\cite{Leinweber:1998uu}.
For momenta smaller than 0.7 GeV, we consider all lattice data. The choice of $p = 0.7$ GeV to distinguish the momenta where the conical and cylindrical cuts
are applied was made to reduce the effects associated to the breaking of rotational symmetry. Indeed, choosing a higher value, let us say 0.8 GeV, then for momenta
below 1 GeV, one can observe different data points for the same $q^2$ that are associated to $D(q^2)$ which differ by much more than a single standard
deviation. In previous studies we have used 1 GeV instead of 0.7 GeV to distinguish the two sets of momenta, but herein given the larger ensembles considered,
to avoid problems due to rotational invariance, this momentum cut has to be lowered. We have checked that the different choice of the scale to
apply the momentum cuts do not change the conclusions reported herein.

The renormalized gluon propagator at $\mu = 3$ GeV can be seen in Fig.~\ref{fig:gluonprops}.
The two ensembles seem to define a unique curve which, again, suggests that finite volume effects are well under control.

As already mentioned, the fitting analysis of the two gluon correlation functions will be performed  using first the tree level Refined Gribov-Zwanziger prediction,
which implements the functional restriction to the first Gribov horizon and defines a local renormalizable quantum field theory \cite{Dudal:2007cw,Dudal:2008sp,Capri:2017bfd}. In order to define a local action, besides the gluon field $A^a_\mu$ the Gribov-Zwanziger action also includes a pair of conjugate bosonic fields
$\left( \overline\varphi^{ac}_\mu  , \varphi^{ac}_\mu  \right)$ and a pair of Grassmann fields $\left( \overline\omega^{ac}_\mu  , \omega^{ac}_\mu  \right)$.
In its original proposal, the perturbative analysis of the action points towards a gluon propagator which vanishes at zero momenta
\begin{equation}
   D(p^2) = \frac{p^2}{p^4 + 2 \, g^2 \, N \, \gamma^4} \ ,
\end{equation}
where $\gamma$ is the Gribov parameter with the dimension of mass, whilst reproducing the original behaviour found by Gribov \cite{Gribov:1977wm}.
This prediction is not compatible with the outcome of contemporary lattice simulations. Though, the predictions of the Gribov-Zwanziger action can be reconciled
with the lattice results by taking into account certain dimension two condensates, associated to local composite operators \cite{Dudal:2007cw,Dudal:2008sp,Gracey:2010cg}.

These condensates show up as minima of the corresponding effective potential \cite{Dudal:2011gd} and introduce further mass scales into the theory which are labeled as
\begin{eqnarray}
  \langle A^a_\mu \, A^a_\mu \rangle & \longrightarrow & -m^2 \, ,\\
  \langle \overline\omega^{ab}_\mu \, \omega^{ab}_\mu \rangle & \longrightarrow & M^2 \, , \\
  \langle \varphi^{ab}_\mu \, \varphi^{ab}_\mu \rangle & \longrightarrow & \rho \,  , \\
  \langle \overline\varphi^{ab}_\mu ~ \overline\varphi^{ab}_\mu \rangle & \longrightarrow & \rho^\dagger \, .
\end{eqnarray}
It follows that by taking into account all the above condensates, the tree level gluon propagator reads
\begin{equation}
   D(p^2) =
   \frac{p^4 + 2 \, M^2 \, p^2 + M^4 - \rho^\dagger \rho}{p^6 + \left(m^2 + 2 \, M^2\right) p^4 + \left( 2 \, m^2 M^2 + M^4 + \lambda^4 - \rho^\dagger \rho\right) p^2  + \iota} \ ,
   \label{EQ:RGZgeneral}
\end{equation}
where $\iota = m^2 \left( M^4 - \rho^\dagger \rho \right) + M^2 \lambda^4  -\frac{\lambda^2}{2} \left( \rho + \rho^\dagger \right)$ and
$\lambda^4 =  2 \, g^2 \, N \, \gamma^4$. In the particular case where $\langle \overline\varphi ~\overline\varphi \rangle = \langle \varphi ~\varphi \rangle$, i.e.
when $\rho$ is a real number, the above expression simplifies to
\begin{equation}
   D(p^2) = \frac{ p^2 + M^2 + \rho}{ p^4 + \left( m^2  + M^2 +  \rho\right) p^2 + m^2 \left( M^2 + \rho \right) + \lambda^4} \ .
   \label{EQ:VRGZ1}
\end{equation}
the case studied first in \cite{Dudal:2010tf}, see also \cite{Cucchieri:2011ig}.
The compatibility of the last functional form, supplemented by an overall residuum $Z$, with the lattice data was investigated in \cite{Dudal:2010tf,Cucchieri:2011ig,Dudal:2012zx}.

Herein, we aim to check how well the lattice data is described by equation (\ref{EQ:RGZgeneral}) using very large ensembles. In order to perform such study, we tried to fit the
lattice data from $p = 0$ up to the momentum $p_{max}$ using the functional forms
\begin{eqnarray}
D_{RGZ} (p^2) & = & Z \, \frac{p^2 + M^2_1}{p^4 + M^2_2 p^2 + M^4_3}  \ ,
   \label{Eq:RGZ} \\
D_{VRGZ} (p^2) & = &  \frac{p^4 + M^2_1 p^2 + M^4_1}{p^6 + M^2_5 p^4 + M^4_4 p^2 + M^6_3}  \ ,
    \label{Eq:VRGZ}
\end{eqnarray}
where $M_i$ are fitting parameters with dimensions of mass. In particular, we are interested to check how far can we go on $p_{max}$,
how stable are the fits and if it is possible to distinguish between the various scenarios. Note that for $D_{VRGZ}(p^2)$ we do not include an overall normalization factor $Z$ or, said otherwise, we set $Z = 1$. This
choice is due to the already ``large number'' of fitting parameters in (\ref{Eq:VRGZ}) and also on the outcome of the fits to the lattice data when
using $D_{RGZ} (p^2)$ -- see discussion below and, in particular, the results reported on Tab.~\ref{tab:rgzfits}, showing that $Z\approx1$ for not too large $p_{max}$.

\section{Results}

\begin{figure}[t] 
   \centering
   \includegraphics[width=3.7in]{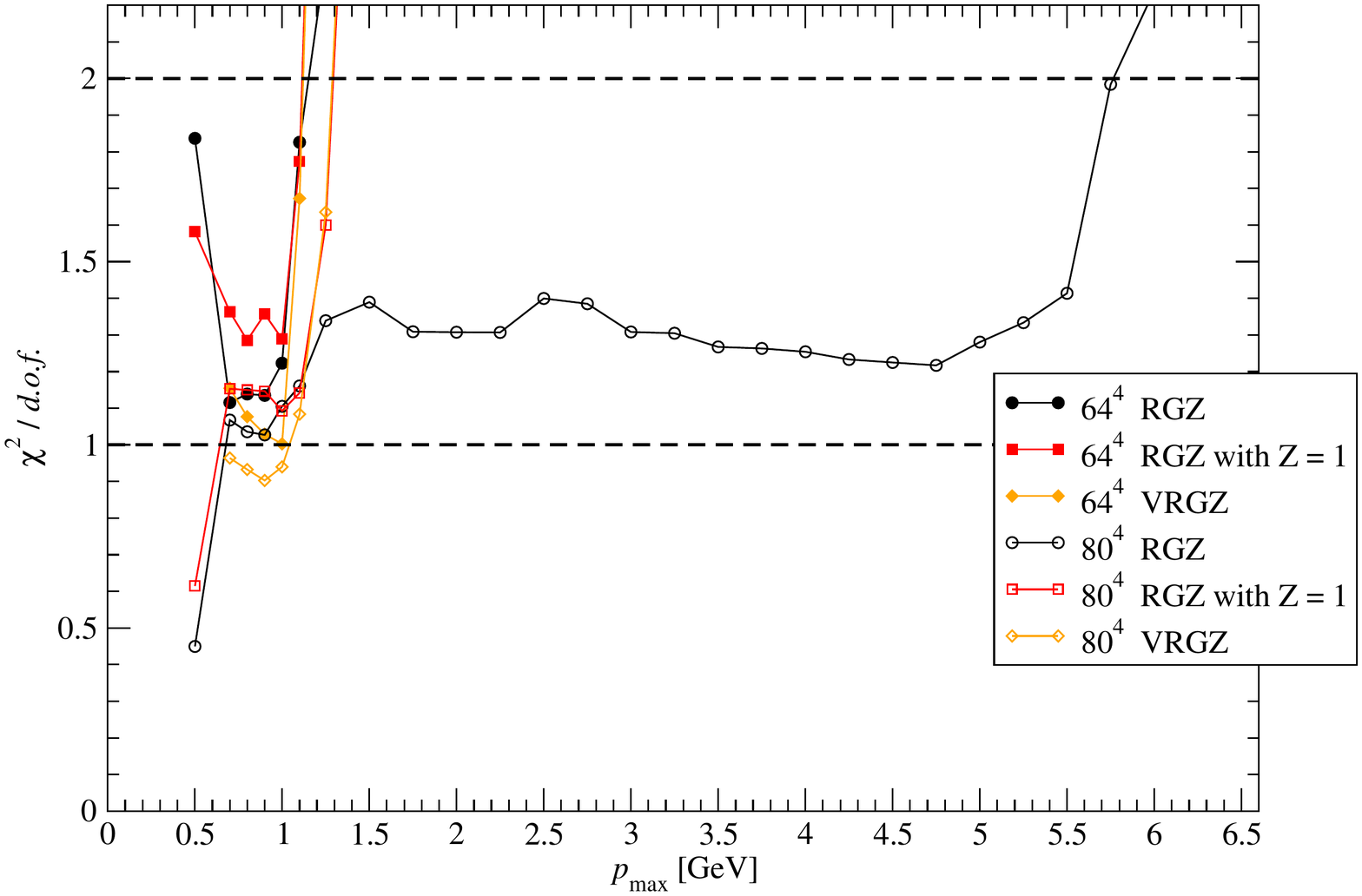}
   \caption{$\chi^2/\text{d.o.f.}$ from fitting the functional forms (\ref{Eq:RGZ}) and (\ref{Eq:VRGZ}) 
                  to the lattice data for $p \in [ 0 , \, p_{max}]$.  Open symbols
                  are associated with the fits to the ensemble having a smaller number of gauge configurations but a larger physical volume,
                  while full symbols are associated with the fits to the larger ensemble but smaller physical volume. }
   \label{fig:chifits}
\end{figure}
For  the functional forms (\ref{Eq:RGZ}) and  (\ref{Eq:VRGZ})  we start fitting the lattice data using $D_{RGZ}(p^2)$ and setting
$p_{max} = 0.5$ GeV. Then,  $p_{max}$ is increased up to 0.7 GeV. For $D_{VRGZ} (p^2)$  the outcome of the fit
with $D_{RGZ}(p^2)$ performed at $p_{max} = 0.7$ GeV is converted into the corresponding picture and we restart fitting the lattice
data with $p_{max} = 0.7$ GeV. The quality of the fits is measured by the value of the $\nu = \chi^2/\text{d.o.f.}$
On Fig.~\ref{fig:chifits} we report on $\nu$ for the fits with a $\chi^2/\text{d.o.f.} \leqslant 2.2$.

As Fig.~\ref{fig:chifits} shows, when one considers the ensemble with the smaller number of gauge configurations, represented by open circles on that
Fig.~\ref{fig:chifits}, the fits to any of the functional forms seem to be compatible with lattice data for a larger range of momenta.
Indeed, for this ensemble it is possible to find a set of parameters with a $\chi^2/\text{d.o.f.} \leqslant 2$ and a $p_{max} \sim 6$ GeV.
On the other hand, for the ensemble with a larger number of gauge configurations ($64^4$ lattice),
which is about four times that of the smaller ensemble ($80^4$ lattice)
and, therefore, has an associated statistical error which is about half of the smaller ensemble, the maximum of $p_{max}$ satisfying the same conditions
is $p_{max} \sim 1$ GeV and, typically, does not go further than 1.5 GeV.

\subsection{Fitting $D_{RGZ}(p^2)$ \label{Sec:RGZ}}
The fits to the Refined Gribov-Zwanziger functional form of Eq. (\ref{Eq:RGZ}) performed using the larger ensemble, i.e.~those to the $64^4$ lattice data,
are compatible with the lattice propagator up to momenta $\lesssim 1$ GeV. On the other hand,
the fits performed with the smaller ensemble, i.e. those relying on the $80^4$ lattice data, show that Eq. (\ref{Eq:RGZ}) is able to reproduce faithfully the
lattice data over a quite larger range of momenta, almost up to the largest momenta available $\sim 6$ GeV.
These results are in agreement with previous studies using ensembles that have, at most, about half of the size of our smaller
ensemble~\cite{Oliveira:2012eh,Duarte:2016iko,Dudal:2010tf,Dudal:2012zx,Cucchieri:2011ig}.
Our conclusion being that the RGZ functional form is compatible with the infrared behaviour of the lattice gluon propagator.

The details of the fits to the RGZ functional form are reported on Tab.~\ref{tab:rgzfits}. In what concerns the fits using the smaller lattice
(larger ensemble), the results for $p_{max} \geqslant 0.7$ GeV are stable and agree well with the corresponding fits for
the larger lattice (smaller ensemble). From the results of fitting the larger ensemble, one concludes that $D_{RGZ}(p^2)$ is compatible
with the lattice data up to $p \sim 1$ GeV. The fits done using the smaller ensemble allow higher values of $p$. However, as can be
seen on Tab.~\ref{tab:rgzfits} from $p = 1.10$ GeV onwards, there is a clear trend visible on the data.

On the other hand, if one sets $Z = 1$, see the lowest part of
Tab.~\ref{tab:rgzfits}, the fitting ranges are quite similar as those found for $D_{RGZ}(p^2)$ and the values
for the various fitting parameters are similar to those found in the upper part of the table, being compatible within one standard deviation.

For comparison we reproduce the results reported in~\cite{Dudal:2010tf}, which also use $\mu = 3$ GeV as renormalization scale,
and where the first study of the compatibility between lattice and RGZ was performed.
For the $64^4$ lattice and $\beta = 6.0$ the authors claimed a $p_{max} = 0.948$ GeV and a
 $Z = 1$, $M^2_1 = 2.579 \pm 0.060$ GeV$^2$, $M^2_2 = 0.536 \pm 0.023$ GeV$^2$ and  $M^4_3 = 0.2828 \pm 0.0052$ GeV$^4$.
 The numbers are compatible within one standard deviation with those quoted on Tab.~\ref{tab:rgzfits} for a similar fitting range.

\begin{table*}[t]
   \centering
   \begin{tabular}{l@{\hspace{0.75cm}} l@{\hspace{0.75cm}} l @{\hspace{0.5cm}} l @{\hspace{0.5cm}}l @{\hspace{0.5cm}} l @{\hspace{0.5cm}} l}
      \hline\hline
       $L$ & $p_{max}$  & $\nu$  & $Z$  & $M^2_1$  & $M^2_2$  & $M^4_3$ \\
       \hline
       64 & 0.50    & 1.84    &  $2.2  \pm 1.3$  &  $0.57  \pm  0.78$   &  $0.421 \pm 0.045$  & $0.14  \pm 0.11$ \\
       64 & 0.70    & 1.12    &  $1.50  \pm 0.17$  &  $1.19  \pm 0.28$    &  $0.417 \pm 0.027$  & $0.200  \pm 0.025$ \\
       64 & 0.80    & 1.14    &  $1.39 \pm 0.17$ &  $1.40   \pm 0.34$     &  $0.432 \pm 0.028$  & $0.216  \pm 0.025$ \\
       64 & 0.90    & 1.14    &  $1.199  \pm 0.084$ & $1.82  \pm 0.25$  &  $0.458 \pm 0.022$  & $0.243  \pm 0.015$  \\
       64 & 1.00    & 1.22    &  $1.088 \pm 0.058$ & $2.16 \pm 0.22$    &  $0.478 \pm 0.021$  & $0.261 \pm 0.013$  \\
       64 & 1.10    & 1.83    &  $0.959 \pm 0.062$ & $2.67 \pm 0.32$    &  $0.511 \pm 0.026$   & $0.285 \pm 0.015$  \\
       \hline
      80 & 0.50     & 0.45     & $2.69  \pm 0.35$   & $0.25\pm 0.13$    & $0.362 \pm 0.012$    & $0.077 \pm 0.028$ \\
      80 & 0.70     & 1.07     & $1.62 \pm 0.19$    & $1.03 \pm 0.28$    & $0.408 \pm 0.033$   & $0.186 \pm 0.029$ \\
      80 & 0.80     & 1.04     & $1.48 \pm 0.17$    & $1.25 \pm 0.30$    & $0.428 \pm 0.033$   & $0.206 \pm 0.027$ \\
      80 & 0.90     & 1.03     & $1.36 \pm 0.13$    & $1.48 \pm 0.29$    & $0.447 \pm 0.030$   & $0.224 \pm 0.022$ \\
      80 & 1.00     & 1.11     & $1.075 \pm 0.087$ & $2.26 \pm 0.34$   & $0.500 \pm 0.029$   & $0.269 \pm 0.018$ \\
      80 & 1.10     & 1.16     & $0.957 \pm 0.066$ & $2.73 \pm 0.34$   & $0.527 \pm 0.029$   & $0.290 \pm 0.016$ \\
      80 & 1.25     & 1.34     & $0.832 \pm 0.062$ & $3.42 \pm 0.44$   & $0.565 \pm 0.031$   & $0.315 \pm 0.016$ \\
      80 & 1.50     & 1.39     & $0.723 \pm 0.037$ & $4.29 \pm 0.37$   & $0.610 \pm 0.026$   & $0.341 \pm 0.012$ \\
      80 & 1.75     & 1.31     & $0.694 \pm 0.018$ & $4.57 \pm 0.22$   & $0.626 \pm 0.018$   & $0.3493 \pm 0.0074$ \\
      80 & 2.00     & 1.31     & $0.697 \pm 0.015$ & $4.54 \pm 0.19$   & $0.624 \pm 0.017$   & $0.3485 \pm 0.0066$ \\
      80 & 2.25     & 1.31     & $0.708 \pm 0.010$ & $4.41 \pm 0.13$   & $0.614 \pm 0.014$   & $0.3441 \pm 0.0051$ \\
      80 & 2.50     & 1.40     & $0.7241 \pm 0.0075$ & $4.22 \pm 0.10$ & $0.598 \pm 0.012$ & $0.3375 \pm 0.0042$ \\
      80 & 2.75     & 1.38    & $0.7288 \pm 0.0063$ & $4.167 \pm 0.087$ & $0.593 \pm 0.011$ & $0.3354 \pm 0.0038$ \\
      80 & 3.00     & 1.31    & $0.7296 \pm 0.0043$ & $4.157 \pm 0.066$ & $0.5922 \pm 0.0097$ & $0.3350 \pm 0.0031$ \\
       \hline\hline
       64 & 0.50    & 1.58     & 1   & $2.31  \pm 0.25$     & $0.452 \pm 0.059$    & $0.257\pm 0.028$ \\
       64 & 0.70    & 1.36     & 1   & $2.472  \pm 0.052$ & $0.494 \pm 0.015$    & $0.2745 \pm 0.0053$ \\
       64 & 0.80    & 1.29     & 1   & $2.469 \pm 0.049$  & $0.494 \pm 0.015$    & $0.2743 \pm 0.0049$ \\
      64 & 0.90     & 1.36    &  1   & $2.515 \pm 0.040$  & $0.506 \pm 0.012$     & $0.2789 \pm 0.0040$ \\
      64 & 1.00     & 1.29    &  1   & $2.521 \pm 0.028$  & $0.5082 \pm 0.0090$ & $0.2795 \pm 0.0027$ \\
      64 & 1.19     & 1.77    &  1   & $2.478 \pm 0.027$  & $0.4955 \pm 0.0089$ & $0.2752 \pm 0.0026$ \\
       \hline
      80 & 0.50      & 0.61    & 1   & $1.96  \pm 0.14$      & $0.378  \pm 0.033$ & $0.220 \pm 0.015$ \\
      80 & 0.70      & 1.15    & 1   & $2.531   \pm 0.064$ & $0.512 \pm 0.019$ & $0.2809 \pm 0.0064$ \\
      80 & 0.80      & 1.15    & 1   & $2.554   \pm 0.060$ & $0.519 \pm 0.018$ & $0.2832  \pm 0.0060$ \\
      80 & 0.90      & 1.15    & 1   & $2.578  \pm 0.056$ & $0.526 \pm 0.017$ & $0.2857 \pm 0.0055$ \\
      80 & 1.00      & 1.09    & 1   & $2.566  \pm 0.044$ & $0.522 \pm 0.014$ & $0.2845 \pm 0.0043$ \\
      80 & 1.10      & 1.14    & 1   & $2.525  \pm 0.036$ & $0.510 \pm 0.011$ & $0.2803  \pm 0.0034$ \\
      80 & 1.25      & 1.60    & 1   & $2.454  \pm 0.035$ & $0.489 \pm 0.011$ & $0.2733 \pm 0.0034$ \\
       \hline\hline
   \end{tabular}
   \caption{Fits to the Refined Gribov-Zwanziger functional form $D_{RGZ}(p^2)$.
   The upper part refers to the fits have $Z$ as a free parameter, while in the lower part of the table $Z = 1$.
    $\nu$ refers to the $\chi^2/\text{d.o.f.}$ For the smaller lattice (larger ensemble) we only show fits with a $\nu < 2$.
   All parameters are in powers of GeV.}
   \label{tab:rgzfits}
\end{table*}

\subsection{Fitting $D_{VRGZ}(p^2)$}
The quality of the description of the lattice gluon propagator by the tree level functional form associated to the Very Refined Gribov-Zwanziger action (VRGZ)
given in Eq.~(\ref{Eq:VRGZ}) can be seen on Fig.~\ref{fig:chifits}. As the Fig.~shows, qualitatively there is no significant
difference between the RGZ and the VRGZ scenarios
and, indeed, the range of $p_{max}$ whose fits have a $\nu = \chi^2/\text{d.o.f.}  \leqslant 2$ is similar. This is somehow unexpected as the number of
fitting parameters in (\ref{Eq:VRGZ}) is larger than in (\ref{Eq:RGZ}) and, therefore, one would expect that the VRGZ would allow for larger values of
$p_{max}$, whilst keeping reasonable values for $\nu$. The main difference will appear at the level of the estimation of the statistical errors, which one expects to be
larger for the scenario with the larger number of fitting parameters.

In Tab.~\ref{tab:vrgzfits} the results of fitting the lattice data to Eq.~(\ref{Eq:VRGZ}) are summarised.
The table includes only the results where the fits have a $\nu = \chi^2/\text{d.o.f.} < 2$.
The  upper part refers to the $64^4$ lattice simulation, while the bottom part  to the $80^4$ lattice simulation.
The excellent agreement between all the fitted parameters and for the two simulations is striking.

All the fits reported on Tab.~\ref{tab:vrgzfits} give a $m^4_1$ and a $m^6_3$ that are always compatible with zero.
If these parameters vanish, then the VRGZ expression nicely reduces to the RGZ functional form
\begin{equation}
D(p^2) =  \frac{p^2 + m^2_2 }{p^4 + m^2_5 p^2 + m^4_4} \  ,
  \label{Eq:VRGZ2RGZ}
\end{equation}
with $\left. m^2_2 \right|_{\mbox{\tiny{VRGZ}}} = \left. m^2_1\right|_{\mbox{\tiny{RGZ}}}$,
$\left. m^2_5 \right|_{\mbox{\tiny{VRGZ}}} = \left. m^2_2 \right|_{\mbox{\tiny{RGZ}}}$,
$\left. m^4_4 \right|_{\mbox{\tiny{VRGZ}}} = \left. m^4_3 \right|_{\mbox{\tiny{RGZ}}}$ and $Z = 1$.

The fits on Tabs.~\ref{tab:rgzfits} and~\ref{tab:vrgzfits} give
\begin{displaymath}
\left. m^2_2 \right|_{\mbox{\tiny{VRGZ}}} = 2.578(60) \mbox{ GeV}^2 \quad\mbox{ and }\quad
\left. m^2_1\right|_{\mbox{\tiny{RGZ}}} = 2.521(28) \mbox{ GeV}^2,
\end{displaymath}
\begin{displaymath}
\left. m^2_5 \right|_{\mbox{\tiny{VRGZ}}} = 0.540(35) \mbox{ GeV}^2 \quad\mbox{ and }\quad
\left. m^2_2 \right|_{\mbox{\tiny{RGZ}}} = 0.5082(90) \mbox{ GeV}^2,
\end{displaymath}
\begin{displaymath}
 \left. m^4_4 \right|_{\mbox{\tiny{VRGZ}}} = 0.290(13) \mbox{ GeV}^4 \quad\mbox{ and }
  \left. m^4_3 \right|_{\mbox{\tiny{RGZ}}} = 0.2795(27) \mbox{ GeV}^4
\end{displaymath}
for the $64^4$ lattice data and for a $p_{max} = 1$ GeV, while for the $80^4$ lattice data
it follows that
\begin{displaymath}
\left. m^2_2 \right|_{\mbox{\tiny{VRGZ}}} = 2.525(36) \mbox{ GeV}^2 \quad\mbox{ and }\quad
\left. m^2_1\right|_{\mbox{\tiny{RGZ}}} = 2.580(64) \mbox{ GeV}^2,
\end{displaymath}
\begin{displaymath}
\left. m^2_5 \right|_{\mbox{\tiny{VRGZ}}} = 0.539(29) \mbox{ GeV}^2 \quad\mbox{ and }\quad
\left. m^2_2 \right|_{\mbox{\tiny{RGZ}}} = 0.522(14) \mbox{ GeV}^2,
\end{displaymath}
\begin{displaymath}
 \left. m^4_4 \right|_{\mbox{\tiny{VRGZ}}} = 0.289(11) \mbox{ GeV}^4 \quad\mbox{ and }
  \left. m^4_3 \right|_{\mbox{\tiny{RGZ}}} = 0.2845(43) \mbox{ GeV}^4
\end{displaymath}
for a $p_{max} = 1.1$ GeV.
The good agreement of the numerics suggests that, indeed, the lattice data in the infrared region defined by $p \lesssim 1$ GeV
is well described by the tree level Refined Gribov-Zwanziger prediction for the gluon propagator.
Furthermore, if the VRGZ reduces to the RGZ, then the equality between the condensates
$\langle \overline\varphi ~\overline\varphi \rangle = \langle \varphi ~\varphi \rangle$ should also hold.

\begin{table*}[t]
   \centering
   \begin{tabular}{l@{\hspace{0.75cm}} l @{\hspace{0.5cm}}l @{\hspace{0.5cm}} l @{\hspace{0.5cm}} l @{\hspace{0.5cm}} l @{\hspace{0.5cm}} l}
      \hline\hline
       $p_{max}$  & $\nu$  & $m^4_1$  & $m^2_2$  & $m^6_3$  & $m^4_4$ & $m^2_5$\\
       \hline
         0.70 & 1.15  &  $0.21 \pm 0.35$    &  $2.78 \pm 0.39$   & $0.0233 \pm 0.039$   & $0.335 \pm 0.088$  &  $0.66  \pm 0.22$ \\
  0.80 & 1.08  &  $0.026 \pm 0.067$  &  $2.55 \pm 0.13$   & $0.0029 \pm 0.0075$  & $0.286 \pm 0.021$  &  $0.527 \pm 0.059$ \\
  0.90 & 1.03  &  $0.083 \pm 0.097$  &  $2.65 \pm 0.11$   & $0.009  \pm 0.011$   & $0.303 \pm 0.024$  &  $0.576 \pm 0.064$ \\
  1.00 & 1.00  &  $0.038 \pm 0.054$  &  $2.578 \pm 0.060$ & $0.0043 \pm 0.0060$  & $0.290 \pm 0.013$  &  $0.540 \pm 0.035$ \\
  1.10 & 1.67  &  $-0.013 \pm 0.037$ &  $2.476 \pm 0.047$ & $-0.0015 \pm 0.0042$ & $0.2730 \pm 0.0091$ & $0.491 \pm 0.026$ \\
 \hline
   0.70 & 0.96  &  $0.11 \pm 0.14$    &  $2.85 \pm 0.26$   & $0.013 \pm 0.016$    & $0.327 \pm 0.043$  &  $0.65 \pm 0.12$ \\
  0.80 & 0.93  &  $0.11 \pm 0.13$    &  $2.86 \pm 0.22$   & $0.013 \pm 0.014$    & $0.327 \pm 0.037$  &  $0.65 \pm 0.11$ \\
  0.90 & 0.90  &  $0.11 \pm 0.11$    &  $2.85 \pm 0.18$   & $0.012 \pm 0.013$    & $0.326 \pm 0.031$  &  $0.651 \pm 0.088$ \\
  1.00 & 0.94  &  $0.061 \pm 0.064$  &  $2.695 \pm 0.095$ & $0.0069 \pm 0.0072$  & $0.304 \pm 0.017$  &  $0.585 \pm 0.048$ \\
  1.10 & 1.08  &  $0.031 \pm 0.045$  &  $2.580 \pm 0.064$ & $0.0035 \pm 0.0051$  & $0.289 \pm 0.011$  &  $0.539 \pm 0.034$ \\
  1.25 & 1.64  &  $0.004 \pm 0.038$  &  $2.465 \pm 0.056$ & $0.0005 \pm 0.0043$  & $0.2748 \pm 0.0097$ & $0.494 \pm 0.029$ \\
        \hline\hline
   \end{tabular}
   \caption{Results for the fits to the Very Refined Gribov-Zwanziger functional form. The upper part refers to the fits to $64^4$ lattice data, while the lower part of the
   table to the fits to the $80^4$ lattice data. All parameters are given in powers of GeV.}
   \label{tab:vrgzfits}
\end{table*}

\section{Global Fits: from Infrared to Ultraviolet \label{Sec:Global}}
From the discussion of the previous section, one can claim that we have now a theoretically motivated expression to describe the infrared lattice gluon propagator.
At high energies one expects to recover the usual perturbative behaviour, whose 1-loop renormalization group (RG) improved results reads, for large $p^2$,
\begin{equation}\label{fitt}
   D(p^2) \propto  \frac{1}{p^2} \left[\ln \left(\frac{p^2}{\Lambda^2_{QCD}}\right)\right]^{\, \gamma_{gl}} \ ,
\end{equation}
where $\gamma_{gl} = -\frac{13}{22}$ is the 1-loop gluon anomalous dimension, $\Lambda_{QCD}$ the renormalization group invariant scale, depending on the chosen renormalization scheme. At 1-loop, we can safely reabsorb the $\mu$-dependent pieces that in principle affect Eq.~\eqref{fitt}, including the coupling constant, into a global rescaling factor.

A good global fitting function should interpolate between these two results (RGZ in the infrared, and the above RG-improved expression in the ultraviolet) and, therefore, we investigate how functional forms
of the type
\begin{equation}
   D(p^2) = Z \, \frac{ p^2 + M^2_1}{p^4+ M^2_2 p^2 + M^4_3} \left[ \omega \,  \ln \left(\frac{p^2 + m^2_{g}(p^2)}{\Lambda^2_{QCD}}\right) + 1 \right]^{\, \gamma_{gl}} \ ,
%
   \label{Eq:globalF}
\end{equation}
where $\omega = 11 N \, \alpha_s ( \mu ) / (12 \pi)$ and $\alpha_s ( \mu )$ is the strong coupling constant defined at the renormalization scale $\mu$,
are able to describe the lattice data for the two ensembles considered herein. In the above expression, to regularise the leading log one has to
introduce a ``log-regularisation mass'' $m^2_{g}(p^2)$ which should become negligible at high momentum in order to recover the
perturbative result.

For now, we do not have any actual analytical estimates of  how the one-loop Refined Gribov-Zwanziger action would affect the ultraviolet log structure. The tree level propagator, with its complex pole structure, will generate a complicated cut structure in the complex $p^2$-plane, with various real and complex branch points \cite{Baulieu:2009ha}.

However, such 1-loop computation would still need to be RG resummed, and in presence of multiple scales, including $p^2$, a proper RG resummation is rather complicated and only amenable to a numerical approach \cite{Einhorn:1983fc,Bando:1992wy,Tissier:2011ey}, again bringing us out of the area of closed functional fitting forms.

We will therefore resort to fairly simple parameterizations of the log. We notice that the expressions $D_{RGZ} (p^2)$ and $D_{VRGZ} (p^2)$ can be put in the form
\begin{displaymath}
    \frac{1}{ p^2 + m^2_{g}(p^2)}
\end{displaymath}
and, therefore, they provide natural definitions for a ``regularizing mass''. In particular, the Refined Gribov-Zwanziger approach will  give
\begin{equation}
  m^2_{g} (p^2) = \frac{ M^4_3 + \left( M^2_2 - M^2_1 \right) p^2}{M^2_1 + p^2} \ . \label{Mass_RGZ}
\end{equation}
The fits performed using this functional form unfortunately give a rather poor $\chi^2/\text{d.o.f.}$
Indeed, as discussed later the fits for the larger ensemble and smaller lattice volume have a $\chi^2/\text{d.o.f.} = 2.98$.

Besides the log-regularizing mass inspired by RGZ, we also considered the following cases
\begin{eqnarray}
  m^2_g(p^2) & = & m^2_0 \, , \label{Mass_Const} \\
  m^2_g(p^2)   & = & \frac{m^4_0}{p^2 + \lambda^2} \, , \qquad\qquad \mbox{(Cornwall)}\label{Mass_Cornwall} \\
  m^2_g(p^2)   & = & \lambda^2_0 + \frac{m^4_0}{p^2 + \lambda^2} \, .  \qquad \mbox{(Corrected Cornwall)}\label{Mass_Cornwall_Modified}
\end{eqnarray}
The first case corresponds to a constant mass \cite{Tissier:2011ey,Serreau:2012cg}, the simplest regularization mass that one can think about,
the second case is a modified version of \cite{Cornwall:1981zr}, see also \cite{Aguilar:2007ie} and \cite{Oliveira:2010xc} where it was used to interpret the lattice gluon propagator.
The last mass function is a further modified version of (\ref{Mass_Cornwall}) that allows better values for the $\chi^2/\text{d.o.f.}$, see hereafter.


For our fits, we took $\Lambda_{QCD} = 0.425$ GeV as the MOM-scheme value estimated in \cite{Dudal:2017kxb} for pure Yang-Mills theory, based on the lattice
works \cite{Boucaud:2008gn,Sternbeck:2010xu,Oliveira:2012eh} and
set $\alpha_s( 3 \mbox{ GeV } ) = 0.3837$, following~\cite{Aguilar:2010gm}.
We call the reader's attention that the entries $\alpha_s$, and therefore $\omega$, as well as $\Lambda_{QCD}$ are fixed parameters, not part of the fitting procedure.

\begin{figure}[t] 
   \centering
   \includegraphics[width=3.7in]{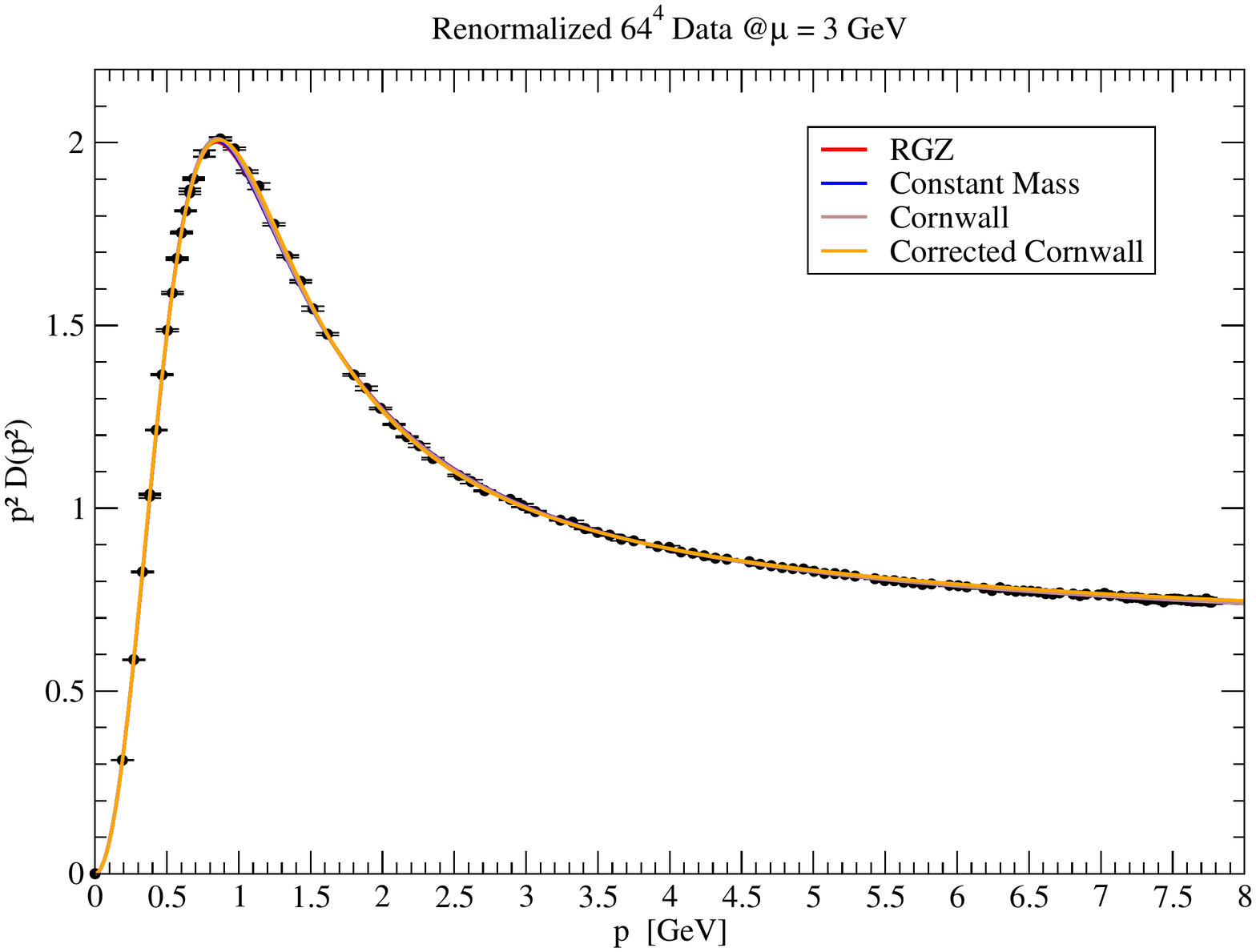} 
   \caption{Gluon dressing function lattice data for the largest ensemble and for simulations on the $64^4$ lattice together with the fits
                with the mutliple regularisation-log running mass functions.}
   \label{fig:globalfits6480}
\end{figure}

\begin{figure}[b] 
   \centering
   \includegraphics[width=3.7in]{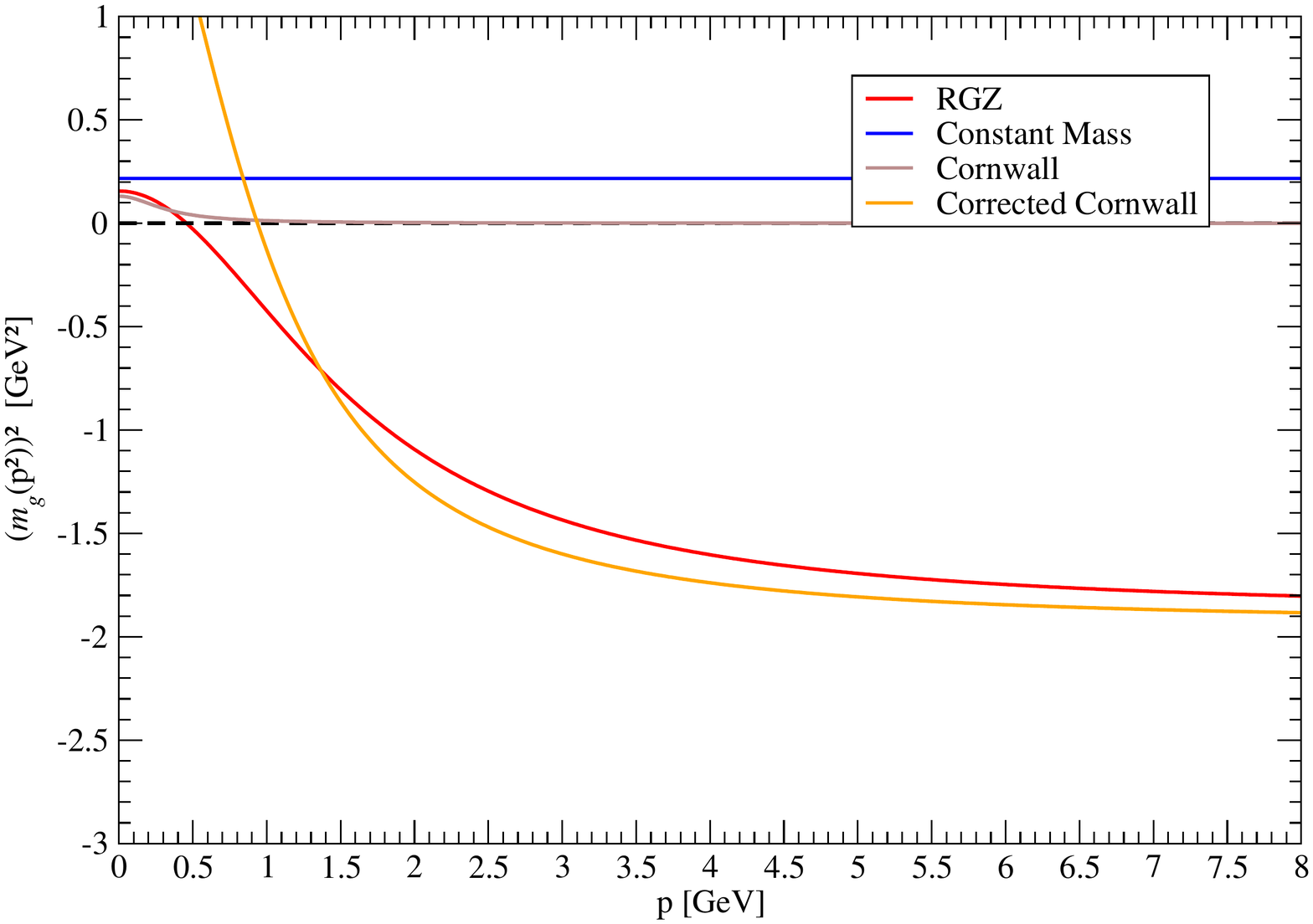} 
   \caption{Regularisation-log running gluon mass for the largest ensemble.}
   \label{fig:runningmass6480}
\end{figure}

The outcome of the global fit to the lattice data using the expressions (\ref{Mass_RGZ})--(\ref{Mass_Cornwall_Modified}) for the log-regularizing mass
can be appreciated on Fig.~\ref{fig:globalfits6480} for the largest ensemble considered herein.
Although for each of the ensembles the curves are almost indistinguishable, in terms of the value of the $\chi^2/\text{d.o.f.}$ there are noticeable differences.

In the following we will show only the fits to the largest ensemble to not overload the text. The outcome of the fits using the smaller but larger physical volume
produce numbers which are compatible with those shown below within one standard deviation.

The fits using the RGZ regularisation mass given by (\ref{Mass_RGZ}) read
\begin{eqnarray}
   Z          & = &  1.36522  \pm 0.00086 \nonumber \\
  M^2_1   & = &  2.518 \pm 0.017 \mbox{ GeV}^2 \nonumber \\
  M^2_2   & = & 0.6379 \pm 0.0070 \mbox{ GeV}^2 \nonumber \\
  M^4_3   & = & 0.3927 \pm 0.0021 \mbox{ GeV}^4 \nonumber
\end{eqnarray}
for a $\chi^2/\text{d.o.f.} = 2.98$.
The fits using a constant mass, see Eq. (\ref{Mass_Const}), give
\begin{eqnarray}
   Z          & = & 1.36486 \pm 0.00097   \nonumber \\
  M^2_1   & = & 2.510 \pm 0.030 \mbox{ GeV}^2 \nonumber \\
  M^2_2   & = & 0.471 \pm  0.014 \mbox{ GeV}^2 \nonumber \\
  M^4_3   & = & 0.3621 \pm 0.0038  \mbox{ GeV}^4 \nonumber \\
  m^2_0   & = & 0.216 \pm 0.026  \mbox{ GeV}^2
\end{eqnarray}
for a $\chi^2/\text{d.o.f.} = 3.15$. The fits using the regularisation mass given by (\ref{Mass_Cornwall}) yields
\begin{eqnarray}
   Z          & = &  1.36636 \pm 0.00079   \nonumber \\
  M^2_1   & = & 2.394 \pm 0.024 \mbox{ GeV}^2 \nonumber \\
  M^2_2   & = & 0.413 \pm 0.013 \mbox{ GeV}^2 \nonumber \\
  M^4_3   & = & 0.3963 \pm 0.0020  \mbox{ GeV}^4 \nonumber \\
  m^2_0   & = &  0.0143 \pm 0.0022 \mbox{ GeV}^4 \nonumber \\
  \lambda^2 & = & 0.109 \pm 0.015  \mbox{ GeV}^2
\end{eqnarray}
for a $\chi^2/\text{d.o.f.} = 2.13$. Finally the fits with a modified Cornwall regularisation mass given by (\ref{Mass_Cornwall_Modified}) give
\begin{eqnarray}
   Z          & = & 1.36992  \pm 0.00072  \nonumber \\
  M^2_1   & = & 2.333 \pm 0.042  \mbox{ GeV}^2 \nonumber \\
  M^2_2   & = &  0.514 \pm 0.024 \mbox{ GeV}^2 \nonumber \\
  M^4_3   & = &  0.2123 \pm 0.0032  \mbox{ GeV}^4 \nonumber \\
  m^2_0   & = & 1.33 \pm 0.13   \mbox{ GeV}^4 \nonumber \\
  \lambda^2 & = &  0.100 \pm 0.035 \mbox{ GeV}^2 \nonumber \\
  \lambda^2_0 & = & -0.954 \pm 0.070 \mbox{ GeV}^2
\end{eqnarray}
for a $\chi^2/\text{d.o.f.} = 1.11$.

For completeness, on Fig.~\ref{fig:runningmass6480} we show the log-regularizing  mass for the various cases studied here. Note
that all the log-regularizing running masses become negative for $p \gtrsim 1$ GeV, except for the case given by  Eq.~\eqref{Mass_Const}.

The global expressions used here  have poles at
\begin{equation}
    p^4 + M^2_2 \, p^2 + M^4_3 = 0
    \label{Eq:IRpoles}
\end{equation}
while branch points are solutions of the equation
\begin{equation}
 0=  p^2 + m^2_g(p^2) = \Lambda^2_{QCD} \, e^{-1/\omega}  = 0.0091978 \dots \mbox{ GeV}^2 \ .
\end{equation}
The first equation defines complex conjugate poles at Euclidean momenta
\begin{equation}
p^2 = - ( 0.20 - 0.32 ) \pm i \, (0.38 - 0.59 ) \mbox{ GeV}^2
\end{equation}
and where typically $| \Im p^2 | \approx | \Re p^2 |$. The branch points associated to the second equation can be on the  real  Euclidean axis
or define a pair of complex conjugate branch points. For the best fits, these
are located at
\begin{equation}
   p^2 = 0.43\pm i \,  1.02 \mbox{ GeV}^2
\end{equation}
for the modified power law mass (\ref{Mass_Cornwall_Modified}). For the power law mass (\ref{Mass_Cornwall}), the branch points are at
\begin{equation}
   p^2 = -0.050 \pm i  \,  0.10 \mbox{ GeV}^2.
\end{equation}
For the constant mass (\ref{Mass_Const}), the branch point is at
\begin{equation}
 p^2 = -0.21 \mbox{ GeV}^2
\end{equation}
and for the RGZ mass (\ref{Mass_RGZ}) at
\begin{equation}
 p^2 = -0.31 \pm i \, 0.34 \mbox{ GeV}^2 \ .
\end{equation}
We call the reader's attention that the conjugate complex poles computed by solving Eq. (\ref{Eq:IRpoles}) using either the outcome of the global fits or the outcome
of the fits up to momenta $p = 1$ GeV give similar numbers.

\section{Summary and Conclusions}

In the current work, we have analysed the compatibility of high statistical lattice calculations of the Landau gauge gluon propagator and the tree
level predictions of the Refined Gribov-Zwanziger (RGZ) action and of the Very Refined Gribov-Zwanziger (VRGZ) action. Compared to previous
studies, the statistical ensembles used here are more than ten times larger.
Our results show that the tree level propagators associated to both actions are
able to describe the lattice data up to momenta $p \lesssim 1$ GeV.
As discussed in the main body of our paper, for smaller statistics, the UV logs are hard to distinguish and as such, the tree level predictions of RGZ extend to much larger momenta. This clearly illustrates the importance of including high enough statistics, as done now, to allow for a trustworthy fit also in the UV.
 Furthermore, we found that the Very Refined Gribov-Zwanziger is
indistinguishable from the RGZ and therefore the relation
$\langle \overline\varphi ~\overline\varphi \rangle = \langle \varphi ~\varphi \rangle$ should hold. This is on par with the earlier observations of \cite{Cucchieri:2011ig}.

From the point of view of the gluon propagator, the analysis performed within the Gribov-Zwanziger approach, associates to the infrared region of the
propagator a pair of complex conjugate poles given by $p^2 = -0.23 \pm i 0.44$ GeV$^2$ from the infrared data, i.e.~the results of Sec.~\ref{Sec:RGZ},
or $p^2 = -0.25 \pm i 0.43$ GeV$^2$ is from considering the global fit of Sec.~\ref{Sec:Global}. Clearly, this result supports that the gluon is not a
physical propagating particle.

For the analysis of the lattice results and within the Gribov-Zwanziger approach, it would be interesting if expressions beyond the tree level approximation were
available, at least for the gluon propagator\footnote{The 1-loop ghost propagator was considered in \cite{Dudal:2012zx,Cucchieri:2016jwg}.}. So far the Gribov-Zwanziger scenario were ``validated'' by the lattice only from studying the pure gauge two point
function correlation functions and relying on its tree level predictions. Despite the difficulties of the calculation suggested,
 it would be interesting to get predictions from higher order
perturbation theory which could be confronted with high precision lattice results and with other non-perturbative approaches.
Recently, in~\cite{Mintz:2017qri} a study of the ghost-gluon vertex within the Gribov-Zwanziger scenario were performed with the results reproducing the lattice data.
However, the poor quality of lattice data available for the three point functions does not allow to arrive at firm conclusions \cite{Cucchieri:2006tf,Cucchieri:2008qm,Maas:2013aia}.
Unfortunately, due to the way lattice simulations access the ghost-gluon or three particle vertices, it will be rather difficult to improve the already published results.

One should also not forget that other approaches than the one analysed here, see for example~\cite{Kondo:2006ih,Shibata:2007eq,Aguilar:2008xm,Gao:2017uox,Tissier:2011ey,Aguilar:2011ux,Kondo:2011ab,Serreau:2012cg,Serreau:2013ila,Pelaez:2014mxa,Frasca:2015yva,Reinosa:2017qtf,Frasca:2017mrh,Siringo:2015gia} and references therein,
are available in the literature and in principle, similar studies as presented in the current paper could be applied to the there suggested analytical forms.
We leave that for a separate study, as the involved functional forms suggested by these other pictures are far more complicated, next to sometimes containing extra fitting parameters, so getting a decent $\chi^2/{d.o.f.}$ might not be so easy. Moreover, to get access to the typical $\ln(p^2)^{-\frac{13}{22}}$ RG  behaviour typically requires a numerical RG approach, making directly fitting a daunting task\footnote{See for example \cite{Tissier:2011ey}. Solutions to the (exact) functional renormalization group are also fully numerical in nature \cite{Cyrol:2016tym,Cyrol:2017ewj}, just as those of the Dyson-Schwinger equations \cite{Aguilar:2008xm,Fischer:2008uz}. }.

\section*{Acknowledgements}
The authors acknowledge the Laboratory for Advanced
Computing at University of Coimbra for providing HPC
computing resources Navigator that have contributed to
the research results reported within this paper (URL
http://www.lca.uc.pt). This work was granted access to
the HPC resources of the PDC Center for High Performance
Computing at the KTH Royal Institute of Technology,
Sweden, made available within the Distributed
European Computing Initiative by the PRACE-2IP, receiving
funding from the European Community's Seventh
Framework Programme (FP7/2007-2013) under grand
agreement no. RI-283493. The use of Lindgren has
been provided under DECI-9 project COIMBRALATT.
We acknowledge that the results of this research have
been achieved using the PRACE-3IP project (FP7 RI312763)
resource Sisu based in Finland at CSC. The use
of Sisu has been provided under DECI-12 project COIMBRALATT2.
P.~J.~Silva acknowledges support by FCT under contracts SFRH/BPD/40998/2007 and
SFRH/BPD/109971/2015. O.~Oliveira and P.~J.~Silva acknowledge financial support from FCT Portugal
under contract with reference UID/FIS/04564/2016.
Part of this work is included in INCT-FNA Proc. No. 464898/2014-5 project. O. Oliveira acknowledge support from CAPES
process 88887.156738/2017-00 and FAPESP grant number 2017/01142-4. D.~Dudal acknowledges financial support from KU
Leuven IF project C14/16/067.


\end{document}